\documentclass[11pt,english]{article}
\usepackage[latin9]{inputenc}
\usepackage{geometry}
\usepackage{verbatim}
\usepackage{textcomp}
\usepackage{amsmath}
\usepackage{amssymb}
\usepackage{esint}

\makeatletter
\@ifundefined{definecolor}
 {\usepackage{color}}{}

\usepackage{babel}

\usepackage{graphicx}

\makeatother

\usepackage{babel}

\begin{document}

\baselineskip 14 pt \parskip 12 pt

\begin{titlepage} 

\begin{flushright}
{\small{}IITH/HEP/XX/17} \\
{\small{}LMU-ASC 23/17}
\par\end{flushright}

\begin{center}
\vspace{2mm}

\par\end{center}

\begin{center}
\textbf{\Large{}Holographic bulk reconstruction beyond (super)gravity}\textbf{ }
\par\end{center}

\begin{center}
Shubho R. Roy$^{1,2}$ and Debajyoti Sarkar$^{3,4}$, \\
 
\par\end{center}

\begin{center}
$^{1}$\textsl{\small{}Department of Physics}\textsl{ }\\
\textsl{\small{}Indian Institute of Technology, Hyderabad, Kandi,
Sangareddy 502285, Medak, Telengana, India}\textsl{ }\\
\texttt{\textsl{\small{}sroy@iith.ac.in}}~\\

\par\end{center}{\small \par}

\begin{center}
$^{2}$\textsl{\small{}Racah Inst. of Physics}\\
\textsl{\small{}Hebrew University of Jerusalem, Jerusalem 91904 Israel}\\

\par\end{center}{\small \par}

\begin{center}
$^{3}$\textsl{\small{}Arnold Sommerfeld Center}\\
\textsl{\small{} Ludwig-Maximilians-University, Theresienstr. 37,
80333 M\"{u}nchen, Germany}\\
\texttt{\small{} debajyoti.sarkar@physik.uni-muenchen.de }\\

\par\end{center}{\small \par}

\begin{center}
$^{4}$\textsl{\small{}Max-Planck-Institut f\"ur Physik (Werner-Heisenberg-Institut)}\\
\textsl{\small{} F\"ohringer Ring 6, D-80805 M\"{u}nchen, Germany}
\par\end{center}{\small \par}

\vskip 0.3 cm
\begin{abstract}
We outline a \emph{holographic} recipe to reconstruct $\alpha'$ corrections to AdS (quantum) gravity from an underlying CFT in the strictly planar limit ($N\rightarrow\infty$). Assuming that the boundary CFT can be solved in principle to all orders of the 't Hooft coupling $\lambda$, for scalar primary operators,
the $\lambda^{-1}$ expansion of the conformal dimensions can be mapped
to higher curvature corrections of the dual bulk scalar field action. Furthermore, for the metric perturbations in the bulk, the AdS/CFT operator-field isomorphism forces these corrections to be of the Lovelock type. We demonstrate this by reconstructing the coefficient of the leading Lovelock correction, aka the Gauss-Bonnet term in a bulk AdS gravity action using the expression of stress-tensor two-point function up to sub-leading order in $\lambda^{-1}$.
\end{abstract}
\end{titlepage}


\section{Introduction\protect
}\label{sec:intro}

In the absence of a fully non-perturbative formulation of (super)string
theory, it is pragmatic to think of the Maldacena duality \cite{Maldacena:1997re,Aharony:1999ti}
as furnishing a manifestly holographic \emph{definition} of quantum
gravity in asymptotically anti de Sitter spaces in terms of a conformal
field theory. This definition is in terms of a large-$N$ conformal
field theory supported on the conformal boundary of the AdS space.
Now we have examples of AdS holography which are neither supersymmetric,
nor require ten or eleven spacetime dimensions. This is the case for e.g. the duality of higher spin gravity in AdS$_{3}$ with $W_{N}$ minimal models in
two dimensions \cite{Gaberdiel:2010pz} and that of (Vasiliev) higher
spin gravity theory \cite{Vasiliev:2003ev} in AdS$_{4}$ with the
$O(N)$ vector model in 2+1 dimensions \cite{Klebanov:2002ja}. Thus
we have come to realize that AdS/CFT is more general than the original
string theory examples where it was first discovered and AdS/CFT can
be elevated to a constructive principle or a starting point/definition
for quantum gravity in asymptotically AdS spaces in terms of CFT degrees
of freedom. Although this definition is manifestly background dependent,
it is completely non-perturbative.%
\footnote{Perhaps, this is the paradigm in constructing quantum gravity in general,
i.e. background dependence (through its asymptotic symmetries) is
an essential ingredient akin to the choice of a global symmetry group
in an ordinary quantum field theory. Perhaps, just as it does not make
much sense to talk about quantum field theories with different (global)
symmetry groups in the same Hilbert space, it does not make sense
to talk of an arbitrarily background \emph{independent} formulation
of quantum gravity. See \cite{Banks:2003vp} for similar arguments. %
} For quantum gravity practitioners, the task then becomes to extract
the quantum gravitational degrees of freedom from the boundary CFT
Hilbert space.%
\footnote{However, the more prevalent use of this duality has been to do with the
exact obverse, i.e. extract CFT observables (correlation functions)
from semiclassical gravity using the GKPW prescription \cite{Gubser:1998bc,Witten:1998qj},
for example applications of AdS/CFT in condensed matter or QCD.%
}\\

In a series of papers \cite{Hamilton:2005ju,Hamilton:2006az,Hamilton:2006fh},
a reformulation of the Lorentzian version of the AdS$_{d+1}$/CFT$_{d}$
correspondence \cite{Balasubramanian:1998sn,Balasubramanian:1999ri}
was worked out in the leading semiclassical (super)gravity approximation,
$N\rightarrow\infty,\lambda\rightarrow\infty$. This reformulation
was based on mapping \emph{normalizable} bulk fields $\Phi(z,x)$
with asymptotic fall-offs, $\Phi(z,x)\overset{z\rightarrow0}{\sim}z^{\Delta}\phi_{0}(x),$
to local CFT operators $\mathcal{O}_{\Delta}(x)$ with scaling dimensions
$\Delta$ \cite{Klebanov:1999tb}. Namely, 
\[
\phi_{0}(x)\leftrightarrow\mathcal{O}_{\Delta}(x).
\]
Here the boundary is located at $z\rightarrow0$ and the boundary coordinates have been collectively denoted by $x$.
The central aim of this reformulation was to recover approximate locality
in the bulk in the most transparent manner - by mapping on-shell bulk
insertions to a delocalized (\emph{smeared}) boundary (CFT) operator
with compact support on the boundary,
\[
\phi(z,x)\text{\ensuremath{\leftrightarrow}}\int dx'K(x'|x,z)\:\mathcal{O}_{\Delta}(x').
\]

This was an improvement over earlier attempts \cite{Banks:1998dd,Bena:1999jv,Balasubramanian:1999ri},
which generally involved representation of a local bulk insertion
in terms of a non-local CFT operator with support over the \emph{entire}
boundary and hence required delicate cancellations to recover bulk
locality at leading order in $1/N$ expansion. The smearing function immediately
reproduces the bulk correlators in terms of the boundary correlators,
for example via
\[
\langle\Phi(x_{1}z_{1})\Phi(x_{2},z_{2})\rangle=\int\: dx'_{1}dx'_{2}\; K(x'_{1}|x_{1},z_{1})\: K(x'_{2}|x_{2},z_{2})\:\langle\mathcal{O}_{\Delta}(x'_{1})\mathcal{O}_{\Delta}(x'_{2})\rangle.
\]

This boundary-to-bulk map or the \emph{smearing function}, \textbf{$K(x^\prime|x,z)$}
has been constructed not just for (spinless) scalar fields, but also
for massive and massless (vector) gauge fields, as well as the spin-two
graviton \cite{Kabat:2012hp} and free higher spin fields \cite{Sarkar:2014dma}. Generically, the smearing functions
are nonvanishing only for points on the boundary which are spacelike
separated%
\footnote{For bulk gauge field insertions, the support is over lightlike separated
points on the boundary.%
} from the local bulk insertion. Further, perturbative quantum gravity
corrections i.e. $1/N$ non-planar effects to the smearing picture
was worked out subsequently \cite{Kabat:2011rz,Kabat:2012av,Kabat:2013wga}.
All this was done in the supergravity (SUGRA) approximation $\lambda\rightarrow\infty$
where the gravity action is just the cosmological Einstein-Hilbert
action. In the original string theory examples, the Regge slope $\alpha'$
is related to the inverse powers of $\lambda$, and such $\alpha'$ corrections
are expected to give rise to ``stringy'' higher derivative corrections
to the (cosmological) Einstein-Hilbert action. However in the
generic case, these $1/\lambda$ corrections i.e. ``stringy'' corrections
to the AdS gravity are yet to be worked out from the CFT. The aim
of this note is to exactly supply that.

The outline of the paper is as follows. In section \ref{sec:mattercorr} and subsection \ref{sec:ls_for_matter}, we introduce massive scalar fields in the bulk interacting with higher curvature terms in a fixed background. We point out how does the concept of a ``string length" emerges from the $\lambda$ corrected anomalous dimensions of the corresponding boundary primary operators. This makes clear the connection between the higher curvature terms in the bulk and the $1/\lambda$ corrections at the boundary. Section \ref{sec:gravcorr} and subsequently subsection \ref{sec:gravcorreom} deal with the higher derivative corrections towards the usual Einstein-Hilbert action. We point out that the equality of the components of the bulk fields and the number of degrees of freedom of the dual operator in the CFT,
forces these corrections to be of the Lovelock type. In section \ref{sec:metricpert} we solve for the metric perturbation equation obtained in the previous section in order to construct the first sub-leading order (in $1/\lambda$ expansion) smearing function for the gravitons. Finally we conclude in section \ref{sec:concl}. The appendix \ref{sec:Gravitational-Perturbation} collects some important formulas.

\section{Matter corrections beyond (super)gravity}\label{sec:mattercorr}

The standard lore in AdS/CFT is that field operators $\Phi(z,x)$
in the quantum gravity side, are described using a bulk (AdS) langrangian
(action) with parameters determined by the anomalous (conformal)
dimension $\Delta$ of the dual operator in the CFT, $\mathcal{O}_{\Delta}(x)$.
In particular, in the planar limit, i.e. $N\rightarrow\infty$ only
connected correlators survive, which generates purely \emph{quadratic}
terms in the bulk (AdS) lagrangian:%
\footnote{Also the strictly planar limit in the CFT implies the vanishing bulk Newton's
constant limit $G_{N}\rightarrow0$, so the matter fields do not back-react
on the AdS geometry. Thus one can safely operate in the \emph{probe}
approximation for matter fields.%
}
\[
\mathcal{L}=-\frac{1}{2}\partial_{M}\Phi\partial^{M}\Phi-\frac{1}{2}m^{2}(\lambda)\Phi^{2}.
\]
When non-planar corrections are included then even in the leading
planar limit, anomalous dimensions of CFT operators receive corrections
from the marginal coupling, $N$ which could be both perturbative
and non-perturbative in nature. Of course, for \emph{supersymmetric}
CFT's there are special cases of BPS-operators which are protected
against such corrections, but since we are considering generic CFT's
or even non-BPS operators in a supersymmetric CFT, we will not restrict ourselves to such special or protected operators. Computing such anomalous
dimensions of gauge invariant operators, at finite values of the coupling,
is \emph{the} fundamental problem in gauge field theories. In general
accomplishing this solution requires fundamentally new insights or
new methods into solving gauge field theories non-perturbatively.
However for the case of \emph{large N} gauge theories, it might be
that the field theory is integrable in the planar limit, for example
as is with the case for $\mathcal{N}=4$ Super Yang Mills theory or ABJ(M) theory \cite{Aharony:2008ug,Aharony:2008gk}.

In a generic large $N$ CFT we have two parameters, a (planar) factorization
parameter $N$ and an exactly marginal coupling $\lambda$. AdS/CFT
isomorphism demands an equality of dimensionless parameters on either
side. In the (super)gravity limit, the emergent AdS spacetime has
two \emph{dimensionfull} parameters, the $AdS_{d+1}$ radius
$R$ and Newton's gravitational constant $G_{N}$. Their ratio constitutes
a single dimensionless parameter which is defined to be
\begin{equation}
\frac{R^{d-1}}{b(d)G_{N}}\equiv N^{2}.\label{eq: AdS/CFT definition of Planck Length}
\end{equation}
Here $b(d)$ is a numerical constant dependent on the spacetime dimensionality.\\

For example in the most extensively explored case of the duality between
type IIB strings on $AdS_{5}\times S^{5}$ and $\mathcal{N}=4$ SYM
\cite{Maldacena:1997re}, the bulk-boundary dictionary between the string coupling $g_s$, string length $l_s$ and Yang-Mills coupling $g_{YM}$ is (omitting \emph{all} numerical factors which are dependent on the spacetime dimensions),
\begin{eqnarray}
g_{s} & = & g_{YM}^{2},\nonumber \\
\left(\frac{R}{l_{s}}\right)^{4} & \sim & Ng_{YM}^{2}=\lambda.\label{eq: TypeIIB/SYM dictionary}
\end{eqnarray}
Also, we have the following relation between Planck length (related to 10-dimensional Newton's constant), string coupling and string length:
\begin{equation}
l_{p}^{8}=g_{s}^{2}l_{s}^{8}\label{eq: Newton's constant in terms of string length}
\end{equation}
Combining (\ref{eq: TypeIIB/SYM dictionary}) and (\ref{eq: Newton's constant in terms of string length}),
we have
\begin{equation}
\frac{R^{8}}{l_{p}^{8}}\sim N^{2}\label{eq: intermediate result}
\end{equation}
Ten dimensional Newton's constant is $l_{p}^{8}$ but after dimensionally
reducing the $S^{5}$ directions, the five dimensional (AdS) Newton's
constant becomes $\tilde{l}_{p}^{3}=l_{p}^{8}/R^{5}.$ Substituting, $l_{p}^{8}=\tilde{l}_{p}^{3}R^{5}$
in (\ref{eq: intermediate result}), we thus obtain, 
\[
\frac{R^{3}}{\tilde{l}_{p}^{3}}\sim N^{2}.
\]
modulo factors depending on dimensionality of the AdS spacetime.\\

The dependence on the coupling of the dual CFT $\lambda$ is much more nontrivial to deduce for an arbitrary CFT, i.e. understanding the second dimensionless parameter or the emergence of a third length scale. However, as in \eqref{eq: TypeIIB/SYM dictionary}, it turns out that in the well-known examples of string theory/CFT dualities, $\lambda$ corresponds to string length (squared) $\alpha'$.
The existence of a dual (super)gravity theory is obtained when the limit 
$\lambda\rightarrow\infty$ is taken. However from the perspective that AdS/CFT is much more general and examples are known which do not require the existence of supersymmetry or ten spacetime dimensions, the equality of ``fundamental''
parameters on the CFT side and gravity side tells us that for each marginal
coupling $\lambda$ there should be a new length scales $l_{s}$
in the bulk, which capture effects of \emph{extended} \emph{classical}
\emph{probes} of the bulk geometry: 
\[
l_{s}^{2}\sim\frac{R^{2}}{\lambda^{\alpha}}.
\]
Here $\alpha$ is determined by the dimensionality of the CFT. In
case of $d=4$, $\alpha=1/2$ as has been derived from methods exploiting
planar integrability - the anomalous dimensions of operators, such
as the Konishi operator receive their first corrections to be of the
order $1/\sqrt{\lambda}$ \cite{Gromov:2009zb}. Since we are reconstructing the bulk/gravity side from the CFT, our work intends to take as an input the expressions of anomalous dimensions of some operator in arbitrary order of $1/\lambda$ (obtained by some pure
CFT method) and as an output determines
the form of the correction terms to be added to the dual bulk field action. However, in practice what we do in the following is to make an educated guess of the bulk correction terms (curvature corrections) and then use the HKLL dictionary to relate the coefficients of these correction terms to the dimensionless coefficients appearing in the expression for scaling dimensions or some other dimensionless coefficients of the correlation functions of the dual boundary operators. Once this relation is established, they uniquely determine the coefficients of the bulk terms in terms of the boundary correction. Thus we are reverse-engineering the CFT data to constrain the correction terms added to the bulk action.\footnote{Note that the smearing functions can also be extracted purely from the boundary data in some simple cases just from symmetry considerations. See e.g. \cite{Verlinde:2015qfa,Miyaji:2015fia,Nakayama:2015mva,Kabat:2017mun,Goto:2017olq}. This is also true in HKLL construction in some sense, where the smearing function is the unique kernel one can write down which will satisfy AdS covariance \cite{Kabat:2012hp,Kabat:2012av,Kabat:2013wga}.}

\subsection{Defining a string length}\label{sec:ls_for_matter}

Here we revisit the case of a bulk scalar, $\Phi\leftrightarrow\mathcal{O}_{\Delta}$.
Duality relates mass $m$ of scalar field in bulk to conformal dimension $\Delta$ of boundary primaries. If the bulk lagrangian is given by 
\begin{equation}
\mathcal{L}=-\frac{1}{2}g^{MN}\partial_{M}\Phi\partial_{N}\Phi-\frac{1}{2}m^{2}(\Delta)\Phi^{2}, \label{bulk free scalar lagrangian}
\end{equation}

then it is well-known that at $N,\lambda\rightarrow\infty$
\[
\Delta_{\infty}=\frac{d}{2}+\sqrt{\frac{d^{2}}{4}+m^{2}R^{2}}.
\]
Stringy corrections can sense the curvature, so on general grounds
including $1/\lambda$ effects, we can write down a modified bulk lagrangian for
the scalar which includes higher curvature corrections:
\begin{eqnarray}
\mathcal{L} & = & -\frac{1}{2}g^{MN}\partial_{M}\Phi\partial_{N}\Phi-\frac{1}{2}m^{2}\Phi^{2}+l_{s}^{2}\left(a\,\mathcal{R}^{MN}\partial_{M}\Phi\partial_{N}\Phi+b\mathcal{R}g^{MN}\partial_{M}\Phi\partial_{N}\Phi+c \,m^2 \mathcal{R}\Phi^{2}\right).\label{eq: Curvature couplings due to 1/lambda corrections}\\
\nonumber 
\end{eqnarray}
Here $\mathcal{R}^{MN}$ and $\mathcal{R}$ are respectively the Ricci tensor and Ricci scalars for the background and $a, b, c$ are constant, dimensionless coefficients which do not depend on any bulk parameters.\footnote{Note that in what follows we are neglecting terms such as $\Phi\nabla \mathcal{R}\nabla\Phi$ for two reasons. First, we contain ourselves with only higher curvature interactions and secondly, due to expansion around pure AdS, for which $\mathcal{R}$ is a constant, such terms drop out. It can also be partially integrated out to give a term going as $\mathcal{R}(\partial\Phi)^2$, which we already considered.} Also in what follows, the capitalized latin indices will denote bulk coordinates and we will reserve the greek indices $\mu,\nu$ etc. to denote the boundary coordinates. Around pure AdS, which is maximally symmetric,
\[
\mathcal{R}_{AB}\propto g_{AB}
\]
and the above modified scalar action simplifies to (on restoring the canonical normalization of the kinetic term)
\begin{equation}
\mathcal{L}=-\frac{1}{2}g^{MN}\partial_{M}\Phi\partial_{N}\Phi-\frac{1}{2}m^{2}\left(1+\bar{c}\frac{l_{s}^{2}}{R^{2}}+O\left(\frac{l_s}{R} \right)^4\right)\Phi^{2},\label{eq: Curvature couplings in one term}
\end{equation}
where $\bar{c}$ is an dimensionless order one constant dependent on the spacetime dimensions and so are the coefficients $a, b, c$ appearing in the higher curvature correction terms in (\ref{eq: Curvature couplings due to 1/lambda corrections}). Thus the overall effect is just a correction of the mass parameter in leading order (infinite $\lambda$) lagrangian (\ref{bulk free scalar lagrangian}).
This then immediately provides the change in the conformal dimension through the asymptotic fall-off 
\begin{equation}
\Delta(\lambda)=\Delta_{\infty}\left(1+\bar{c}f(\Delta_{\infty})\frac{l_{s}^{2}}{R^{2}}+\ldots\right).\label{eq: Bulk fall-off modification}
\end{equation}
where $f(\Delta_\infty)$ is a function of only $\Delta_\infty$ and is given by
\[
f(\Delta_\infty)=\frac{\left(\Delta_\infty-\frac{d}{2}\right)^2-\frac{d^2}{4}}{\Delta_{\infty}\left(\Delta_\infty-\frac{d}{2}\right)}.
\]
Note that it goes to an order one constant as well, when we consider conformal primaries with large operator dimensions. In principle, this is what one expects by directly working with the CFT itself, namely, one can compute the conformal dimension as
\begin{equation}
\Delta(\lambda)=\Delta_{\infty}\left(1+O\left(\frac{1}{\lambda^{\alpha}}\right)\right).\label{eq: Conformal dimension directly from CFT}
\end{equation} 
This is for example what was done for the Konishi operator \cite{Gromov:2009zb}.
This is of course the holy grail of field theorists, to solve the
spectral dimension for arbitrary coupling, $\lambda$. However as
quantum gravity practitioners, we will assume that the CFT has been
solved exactly and the spectral dimensions are known to all orders
in $1/\lambda$. Comparing the two expressions for the conformal dimensions
(\ref{eq: Bulk fall-off modification}) and (\ref{eq: Conformal dimension directly from CFT}), we can identify ``string length''
\begin{equation}
\frac{l_{s}^{2}}{R^{2}}\equiv\frac{1}{\lambda^{\alpha}}.\label{eq: String Length in terms of CFT scaling dimensions flow and AdS Radius}
\end{equation}
Thus, finally the smearing function in this case is simply modified to (keeping the SUGRA and boundary normalizations same as in \cite{Hamilton:2006fh})
\begin{equation}\label{eq:bulkhcsmear}
\Phi(t,x,z)= \frac{\Gamma\left[\Delta(\lambda)-\frac{d}{2}+1\right]}{\pi^{d/2}\Gamma\left[\Delta(\lambda)-d+1\right]}\int_{t'^2+y'^2< z^2}dt'd^{d-1}y'\, (2\sigma z')^{\Delta(\lambda)-d}\mathcal{O}_{\Delta(\lambda)}(t+t',x+iy'),
\end{equation}
for AdS covariant bulk-boundary distance 
\begin{eqnarray*}
\sigma(z,x|z^\prime,x^\prime)=\frac{z^2+{z^\prime}^2+(x-x^\prime)^2}{2zz^\prime}
\end{eqnarray*}
in e.g. Poincar\'{e} AdS.

As we mentioned before, at this point one should treat the above smearing function as the prescription to reverse-engineer the bulk fields and correlators from their boundary counterparts. 

\section{Gravity action from $\lambda^{-1}$ corrections in CFT\protect
}\label{sec:gravcorr}

After getting an intuition behind the equivalence between the boundary $\lambda^{-1}$ corrections and bulk higher curvature corrections, we now directly consider modified gravity actions in the bulk. Here we will neglect contributions from any other matter fields and set the stage for computing metric perturbation and hence the modified graviton smearing function in the next sections.

The $\lambda$-corrections in the bulk are better not thought
of as quantum corrections, but \emph{classical} non-locality induced
contributions due to \emph{extended probes of geometry}. For example
in the $\mathcal{N}=4$ SYM/ type IIB case, these are \emph{classical
stringy non-local} effects. Such effects are manifested
in local lagrangian field theory by an infinite number of higher derivative
terms. Thus, one needs to turn them on in the gravity
action to precisely capture the non-localities arising out of
extended probes in the bulk
\begin{equation}
I_{bulk}=\frac{1}{16\pi G_{N}}\int d^{d+1}x\:\sqrt{-g}\left(\mathcal{R}-2\Lambda+\alpha_{1}\mathcal{R}^{2}+\alpha_{2}\mathcal{R}_{\mu\nu}^{2}+\alpha_{3}\mathcal{R}_{\mu\nu\rho\sigma}^{2}+\alpha_{4}\square \mathcal{R}+\ldots\right).\label{eq: lambda corrected action}
\end{equation}
In general the parameters in the bulk action $\Lambda$ and $\alpha_{i}$'s
are functions of the gauge theory parameters $N$ and $\lambda$\,:\footnote{We will later see how to interpret $\alpha_i$'s as a purely boundary quantity without making any reference to AdS radius $R$.}
\begin{eqnarray}\label{eq:alphalambdareln}
\frac{\alpha_{i}}{R^2} & \sim & \frac{1}{\lambda^{\alpha}}+O\left(\frac{1}{\lambda^{2\alpha}}\right),\nonumber\\
\Lambda R^2 & \sim & \Lambda_{0}R^2+\frac{a}{\lambda^{\alpha}}+O\left(\frac{1}{\lambda^{2\alpha}}\right),\quad\mbox{with}\quad\Lambda_{0} R^2 =-\frac{d(d-1)}{2}.
\end{eqnarray}
These dependences are not arbitrary but constrained by the following
principles:
\begin{itemize}
\item The new higher derivative gravity action admits an \emph{exact}
pure AdS solution. This is because the symmetries should remain intact
on both the AdS side and CFT side. On the CFT side, since the conformal
symmetry of the vacuum does not get overhauled by the $\lambda^{-1}$ corrections,
hence a pure AdS space must be a solution to the $\lambda^{-1}$ corrected
bulk field equations as well.
\item This pure AdS solution to the field equations of higher derivative
gravity with the cosmological constant $\Lambda$, is identical to
AdS space with radius $R$ which is a solution to the (two derivative)
Einstein's field equations with cosmological constant $\Lambda_{0}=-\frac{d(d-1)}{2R^{2}}$.
This is necessary so that the definition of Newton's constant (\ref{eq: AdS/CFT definition of Planck Length})
does not get revised/renormalized.%
\footnote{If we allow the AdS radius to change with $\lambda$, then (\ref{eq: AdS/CFT definition of Planck Length})
implies that Planck length needs to change identically, so as to keep
their ratio fixed and equal to $N^{2}$. But change in Planck length
implies that they are \emph{quantum/loop} corrections in the bulk, not
classical string corrections. %
} In the $\mathcal{N}=4$ SYM/ type IIB duality, it is known that
the AdS radius does not get renormalized by stringy $\alpha'$ corrections
owing to supersymmetry \cite{Mazzucato:2009fv}. This demand then
implies \cite{Fukuma:2001uf}:
\end{itemize}
\begin{equation}
\Lambda=\Lambda_{0}+\frac{d(d-3)}{2R^{4}}\left[d(d+1)\alpha_{1}+d\alpha_{2}+2\alpha_{3}\right],\quad\mbox{with}\quad\Lambda_{0}=-\frac{d(d-1)}{2R^{2}}.\label{eq: Fukuma relation}
\end{equation}

\begin{itemize}
\item The $\lambda^{-1}$ corrections in the bulk, being classical corrections
due to non-locality/extended probes, must admit consistent semiclassical
quantizations of gravity (about AdS space). In particular these corrections
must not change the number of (on-shell) degrees of freedom associated
with the graviton as AdS/CFT demands that the number of degrees of
freedom of the graviton must be same as that of the CFT stress
tensor. This is directly manifest in \emph{holographic} gauge of
\cite{Kabat:2012hp}, where there is a \emph{direct} isomorphism
between components of CFT stress-tensor and local graviton insertions
in AdS:
\begin{equation}
h_{\mu\nu}\leftrightarrow T_{\mu\nu}\quad\mbox{with}\quad h_{\mu z}=h_{zz}=0.\label{eq: graviton is isomorphic to Stress tensor}
\end{equation}
Generically, higher derivative actions modify the degrees of freedom
due to the presence of higher derivatives. Isomorphism with gauge
theory/CFT forces the requirement of keeping intact the number of
graviton on-shell degrees of freedom, i.e. forces us to choose a very
particular form of the $\lambda$-corrected action, one which was
obtained in \cite{Zwiebach:1985uq,Zumino:1985dp}:
\[
\alpha_{1}=\alpha_{3}=-\frac{1}{4}\alpha_{2}.
\]
$\alpha_{4}$ can be consistently set to zero as it is the coefficient
of a total derivative term. This term is called a Gauss-Bonnet term \cite{Lovelock:1971yv}
and the specialty of such a term is that the resulting equations of motion contain only second derivatives. For higher order $\lambda^{-1}$ corrections one generates/adds higher order Lovelock terms \cite{Lovelock:1971yv} in the bulk AdS action. Note that if the stress-tensor of the CFT is itself Lorentz anomalous, then of course one can consider general higher derivative terms which are not Lovelock. As an example consider in AdS$_3$, a TMG term (topologically massive gravity term). It corresponds to an extra stress tensor (generally called the anomalous stress tensor) in the dual chiral log CFT.\footnote{We are thankful to Arpan Bhattacharyya for pointing out this example to us.}
\item The stress tensor two-point function (and its higher order functions
in general) determine the coefficient $\alpha_{1}$, since in the
boundary limit
\begin{equation}
\lim_{z,z'\rightarrow0}\langle h_{\mu\nu}(x,z)\: h_{\rho\sigma}(x',z')\rangle_{(SU)GRA}=z^{d-2}z'^{d-2}\langle T_{\mu\nu}(x)T_{\rho\sigma}(x')\rangle_{(S)CFT},\label{eq: Boundary limit holography}
\end{equation}
as was done in \cite{Kabat:2012hp}. The left hand side graviton two-point
function in the Gauss-Bonnet gravity is a function of $\alpha_{1}$.
But there is one subtlety to note here: usually the normalization convention in the  left hand i.e. the (super)gravity side is different from the right hand or the S(CFT) side. For (super)gravity the loop expansion parameter is $\kappa^{2}=16\pi G_{N}$ and the graviton is defined as a perturbation $h_{\mu\nu}$ around some background $g_{\mu\nu}^{(0)}$ like
\[
g_{\mu\nu}=g_{\mu\nu}^{(0)}+\kappa h_{\mu\nu},
\]
and as a result the Newton's constant \emph{does not} show up in
the SUGRA two-point function for gravitons. In the Large $N$ CFT
side however, two-point functions, particularly of boundary stress tensors (and in general all connected correlators) are usually taken to have a norm which scales as $N^{2}$, 
\[
\langle\mathcal{O}_{\Delta}(x)\mathcal{O}_{\Delta}(x')\rangle=\frac{C_{\mathcal{O}}}{\left(x-x'\right)^{2\Delta}},\quad\mbox{with}\quad C_{\mathcal{O}}\sim N^{2}.
\]
The usual practice in a large $N$ CFT is to set the norm of the two-point
function to unity, by defining 
\[
\mathcal{O}\rightarrow\mathcal{O}/\sqrt{C_{\mathcal{O}}}.
\]
But the stress-tensor two-point function is not normalized to unity.
In fact, the norm gives the \emph{central charge}, which is a characteristic
of the field theory (heuristically speaking, it is an indicator
of the field content of the (S)CFT). For an (S)CFT one has \cite{Fradkin:1996is,Erdmenger:1996yc},
\begin{equation}
\langle T_{\mu\nu}(x)T_{\rho\sigma}(x')\rangle=C(N^{2},\lambda)\frac{I_{\mu\nu;\rho\sigma}}{(x-x')^{2d}}.\label{eq: Central Charge as the leading singularity of the two point function of stress tensor}
\end{equation}
Here $I_{\mu\nu,\rho\sigma}$ is some (universal) conformally covariant
structure depending on $(x-x')$ independent of the CFT field content.
The central charge, has an asymptotic expansion about $\lambda\rightarrow\infty$,\footnote{Note that the $1/\lambda$ order correction to the graviton two point function can be absorbed in a redefinition of the Newton's constant $G_N$, instead of the running of the central charge $C$. (We should not confuse this renormalization of $G_N$ with $1/N$ i.e. quantum gravity loop corrections). However, when we go to the graviton three point function level, there are three ``central charges'' corresponding to the coefficients of three distinct tensor structures allowed for the three point function. Thus at the three point level one cannot just absorb the $1/\lambda$ running of three distinct central charges into a single Newton's constant. We thank Dan Kabat for pointing this out to us.} 
\[
C(N^{2},\lambda)=C(N^{2})\: f(\lambda),\qquad f(\lambda)=1+O\left(\lambda^{-\alpha}\right).
\]
(of course recall that we are talking in the strictly planar limit,
otherwise the central charge would, in addition, have terms sub-leading
in $1/N$). Thus we need to make the identification (\ref{eq: graviton is isomorphic to Stress tensor})
in a way so that the boundary limit of the AdS graviton two-point
function reproduces CFT stress tensor two-point function with the
normalization $\sim N^{2}$ (central charge). This is easily achieved
by the GKPW recipe \cite{Gubser:1998bc,Witten:1998qj} of extracting
boundary stress-tensor correlators 
or perhaps more directly for us, through the BDHM (extrapolate) or HKLL dictionary
of AdS/CFT in the \emph{holographic gauge}, where we arrive at the boundary
CFT stress-tensor two point functions, by taking the boundary limit
of the bulk graviton two-point function 
\[
\langle T_{\mu\nu}(x)T_{\lambda\rho}(z,y)\rangle_{CFT}=\lim_{z\rightarrow0}z^{2d}\langle h_{\mu\nu}(z,x)h_{\lambda\rho}(z,y)\rangle_{SUGRA}.
\]
One \emph{always} identifies the central charge in the supergravity
limit ($N,\lambda\rightarrow\infty$) in terms of SUGRA parameters
\begin{equation}
C(N^{2})=n(d)\frac{R^{d-1}}{16\pi G_{N}},\label{eq: Central Charge from GKPW}
\end{equation}
where $n(d)$ is a numerical factor depending on the dimensionality
of the field theory.
\end{itemize}

%

\subsection{CFT induced (four) derivative corrections to bulk action (Gauss-Bonnet)\protect
}\label{sec:gravcorreom}

We now perform the required calculations in order to justify various points that we discussed above. The variation of the action (\ref{eq: lambda corrected action}) restricted to first order in $\lambda^{-\alpha}$ %
gives the field equations for four derivative corrected gravity \cite{Deser:2011xc},\footnote{Recall that for non-compact spaces such as AdS, all total derivative
terms vanish and there is no need to add boundary terms such as Gibbons-Hawking
term.%
} 
\begin{align}
0=&\frac{g_{AB}}{2}\left(\mathcal{R}-2\Lambda+\alpha_{1}\mathcal{R}^{2}+\alpha_{2}Ric^{2}+\alpha_{3}Rie^{2}\right)-\mathcal{R}_{AB}+2\alpha_{1}\left(\nabla_{A}\nabla_{B}\mathcal{R}-g_{AB}\square \mathcal{R}-\mathcal{R}_{AB}\mathcal{R}\right)\nonumber \\
&+\alpha_{2}\left(\nabla^{C}\nabla_{A}\mathcal{R}_{BC}+\nabla^{C}\nabla_{B}\mathcal{R}_{AC}-\square \mathcal{R}_{AB}-g_{AB}\nabla_{C}\nabla_{D}\mathcal{R}^{CD}-2\mathcal{R}_{AC}\mathcal{R}_{B}\,^{C}\right)\qquad\nonumber \\
&-2\alpha_{3}\left(\mathcal{R}_{A}\,^{LMN}\mathcal{R}_{BLMN}+\nabla^{L}\nabla^{M}\mathcal{R}_{ALBM}+\nabla^{M}\nabla^{L}\mathcal{R}_{ALBM}\right).\label{eq:Penultimate form of Field Equation}
\end{align}
Obviously, $Ric$ and $Rie$ signify the Ricci and Riemann tensors for the background. Next, using the (Bianchi) identities,
\[
\nabla^{A}\mathcal{R}_{AB}=\frac{1}{2}\nabla_{B}\mathcal{R},
\]
\[
\nabla^{A}\nabla^{B}\mathcal{R}_{AB}=\frac{1}{2}\square \mathcal{R}\quad\mbox{and}
\]
\[
\nabla^{L}\nabla^{M}\mathcal{R}_{ALBM}=\square \mathcal{R}_{AB}-\frac{1}{2}\nabla_{A}\nabla_{B}\mathcal{R}-\mathcal{R}_{AC}\mathcal{R}_{B}\,^{D}+\mathcal{R}_{ACBD}\mathcal{R}^{CD},
\]
we can recast the field equations in a nicer form
\begin{align}
0=&\mathcal{R}_{AB}-\frac{1}{2}g_{AB}\left(\mathcal{R}-2\Lambda+\alpha_{1}\mathcal{R}^{2}+\alpha_{2}Ric^{2}+\alpha_{3}Rie^{2}\right)+2\alpha_{1}\mathcal{R}\mathcal{R}_{AB}-4\alpha_{3}\mathcal{R}_{AC}\mathcal{R}_{B}\,^{C}\nonumber \\
&+(2\alpha_{2}+4\alpha_{3})\mathcal{R}_{ACBD}\mathcal{R}^{CD}+2\alpha_{3}\mathcal{R}_{ALMN}\mathcal{R}_{B}\,^{LMN}+\left(2\alpha_{1}+\frac{\alpha_{2}}{2}\right)g_{AB}\square \mathcal{R}\nonumber\\
&-\left(2\alpha_{1}+\alpha_{2}+2\alpha_{3}\right)\nabla_{A}\nabla_{B}\mathcal{R}+\left(\alpha_{2}+4\alpha_{3}\right)\square \mathcal{R}_{AB}.\label{eq: Final form of higher derivative Field Equations}
\end{align}
Since we must have AdS spacetime as a solution to these four derivative
gravity field equation, we plug in an AdS space ansatz with the AdS
radius $R$,%
\footnote{This radius is unchanged while the ``bare'' parameters in the lagrangian
are changed as higher and higher derivatives terms are added.%
}
\[
\mathcal{R}_{ABCD}=-\frac{1}{R^{2}}\left(g_{AC}g_{BD}-g_{AD}g_{BC}\right).
\]
The derivative terms become derivatives of the metric and vanish,
and the non-vanishing contribution to the field equations are:
\begin{align*}
\mathcal{R}_{AB}&-\frac{1}{2}g_{AB}\left(\mathcal{R}-2\Lambda+\alpha_{1}\mathcal{R}^{2}+\alpha_{2}Ric^{2}+\alpha_{3}Rie^{2}\right)+2\alpha_{1}\mathcal{R}\mathcal{R}_{AB}-4\alpha_{3}\mathcal{R}_{AC}\mathcal{R}_{B}\,^{C}\nonumber\\
&+(2\alpha_{2}+4\alpha_{3})\mathcal{R}_{ACBD}\mathcal{R}^{CD}+2\alpha_{3}\mathcal{R}_{ALMN}\mathcal{R}_{B}\,^{LMN} =0.
\end{align*}
After contracting with $g^{AB}$ on both sides, we get the \emph{revised}
cosmological constant parameter in the lagrangian,
\begin{equation}
\Lambda=-\frac{d(d-1)}{2R^{2}}+\frac{d(d-3)}{2R^{4}}\left(d(d+1)\alpha_{1}+d\alpha_{2}+2\alpha_{3}\right).\label{eq: Cosmological constant parameter}
\end{equation}

In general, for the higher derivative contributions to vanish one
needs to arrange the $\alpha_i$'s to set the coefficients of the three \emph{independent}
higher derivative terms $\square \mathcal{R},$ $\nabla_{A}\nabla_{B}\mathcal{R}
$ and $\square \mathcal{R}_{AB}$ in the field equations (\ref{eq: Final form of higher derivative Field Equations}) to vanish i.e.
\[
2\alpha_{1}+\frac{1}{2}\alpha_{2}=0,
\]
\[
2\alpha_{1}+\alpha_{2}+2\alpha_{3}=0,
\]
\[
\alpha_{2}+4\alpha_{3}=0.
\]
Evidently this is achieved for \emph{arbitrary} spacetime dimensions
for the Gauss-Bonnet combination i.e.,
\[
\alpha_{2}=-4\alpha_{1};\alpha_{3}=\alpha_{1}.
\]
For this case the field equations take the form,
\begin{eqnarray}
\mathcal{R}_{AB}-\frac{1}{2}g_{AB}\left[\mathcal{R}-2\Lambda+\alpha_{1}\left(\mathcal{R}^{2}-4Ric^{2}+Rie^{2}\right)\right]\qquad\qquad\qquad\qquad\qquad\qquad\qquad\nonumber \\
+2\alpha_{1}\left[\mathcal{R}\mathcal{R}_{AB}-2\mathcal{R}_{AC}\mathcal{R}_{B}\,^{C}-2\mathcal{R}_{ACBD}\mathcal{R}^{CD}+\mathcal{R}_{ALMN}\mathcal{R}_{B}\,^{LMN}\right] & = & 0.\label{eq:Einstein Field Equations in Gauss-Bonnet Gravity}
\end{eqnarray}
Note that the cosmological constant for (asymptotically) AdS backgrounds
in Gauss-Bonnet gravity becomes,
\begin{equation}
\Lambda=-\frac{d(d-1)}{2R^{2}}+\frac{d(d-1)(d-2)(d-3)}{2R^{4}}\alpha_{1}.\label{eq: cosmological constant for Gauss-Bonnet AdS gravity}
\end{equation}
Below, following \cite{D'Hoker:1999jc,Kabat:2012hp} we shall use a slightly
more convenient form of the field equations,
\begin{align}
\mathcal{R}_{AB}&+\frac{d}{R^{2}}\left(1-\frac{\alpha_{1}}{R^{2}}(d-2)(d-3)\right)g_{AB}-\frac{\alpha_{1}}{d-1}g_{AB}\left(\mathcal{R}^{2}-4Ric^{2}+Rie^{2}\right)\nonumber \\
&+2\alpha_{1}\left(\mathcal{R}\mathcal{R}_{AB}-2\mathcal{R}_{AC}\mathcal{R}_{B}\,^{C}-2\mathcal{R}_{ACBD}\mathcal{R}^{CD}+\mathcal{R}_{ALMN}\mathcal{R}_{B}\,^{LMN}\right) = 0.\label{eq:Ricci Equation for Gauss-Bonnet in AdS}
\end{align}

\section{Corrections to bulk (SU)GRA from $1/\lambda$ running of CFT stress tensor correlations\protect
}\label{sec:metricpert}

In this section we finally turn to linearizing the above equation of motion in order to obtain the smearing function to the first sub-leading order in $\lambda^{-1}$ expansion. The graviton equation of motion is obtained by linearizing the field
equations (\ref{eq:Einstein Field Equations in Gauss-Bonnet Gravity}), (\ref{eq:Ricci Equation for Gauss-Bonnet in AdS}),
around the pure AdS solution,
\[
g_{AB}=g_{AB}^{(0)}+h_{AB},
\]
using the following linearized forms around AdS space,
\[
\mathcal{R}^{(1)A}\,_{LMN}=\frac{1}{2}\left(\nabla_{M}\nabla_{N}h_{L}^{A}+\nabla_{M}\nabla_{L}h_{N}^{A}-\nabla_{M}\nabla^{A}h_{LN}-M\leftrightarrow N\right),
\]
\[
\mathcal{R}_{AB}^{(1)}=\nabla_{(A}\nabla^{C}h_{B)C}-\frac{1}{2}\square h_{AB}-\frac{1}{2}\nabla_{B}\nabla_{A}h-\frac{d+1}{R^{2}}h_{AB}+\frac{1}{R^{2}}g_{AB}^{(0)}h,
\]
\[
\mathcal{R}^{(1)}=\nabla^{C}\nabla^{D}h_{CD}-\square h+\frac{d}{R^{2}}h.
\]
%

At the end of the day, quite expectedly, the linearized Ricci equation remains the same,%
\begin{equation}
\left(1-\frac{2(d-2)(d-3)}{R^{2}}\alpha_{1}\right)\left(\nabla_{(A}\nabla^{C}h_{B)C}-\frac{1}{2}\square h_{AB}-\frac{1}{2}\nabla_{B}\nabla_{A}h-\frac{1}{R^{2}}h_{AB}+\frac{1}{R^{2}}g_{AB}^{(0)}h\right)=0,\label{eq: Ricci Equation for Gauss-Bonnet in AdS}
\end{equation}
i.e. identical to that of (super)gravity \cite{D'Hoker:1999jc}, except
for the appearance of an \emph{overall} coefficient which doesn't
alter anything. However, the bulk Green's function (e.g. Feynman or retarded/advanced ones) for the metric
perturbations $G_{AB}(x-y)$ will be affected, 
\begin{eqnarray*}
\left(1-2\alpha_{1}\frac{(d-2)(d-3)}{R^{2}}\right)\left(\nabla_{x^{(A}}\nabla^{x^{C}}G_{x^{B)}x^{C}}(x-y)-\frac{1}{2}\square_{x}G_{AB}(x-y)\right.\\
-\frac{1}{2}\nabla_{x^{B}}\nabla_{X^{A}}G(x-y)\left.-\frac{1}{R^{2}}G_{AB}(x-y)+\frac{1}{R^{2}}g_{AB}^{(0)}G(x-y)\right) & = & \frac{1}{\sqrt{g}}g_{AB}\delta^{d+1}(x-y)
\end{eqnarray*}
because the right hand side is a delta function. In particular the Green's function
will get rescaled by this pre-factor,
\[
G_{GB}\sim\frac{1}{\left(1-2\alpha_{1}\frac{(d-2)(d-3)}{R^{2}}\right)}G_{EH},
\]
where the subscript $GB$ denotes Gauss-Bonnet theory in the bulk
while the subscript denotes Einstein-Hilbert in the bulk. This then
implies that the smearing functions obtained in \cite{Kabat:2012hp}
from the spacelike supported Green's function will also get rescaled
by the $\alpha_{1}$-dependent normalization factor.\footnote{This can also be expected in other intuitive ways. For example, we can derive \eqref{eq: Ricci Equation for Gauss-Bonnet in AdS} by redefining the initial metric perturbation $h_{AB}$, which has the Einstein-Hilbert form, upon absorbing the extra $\alpha_1$ dependent factor in it.} Hence, following \cite{Kabat:2012hp}, the final graviton smearing expression becomes ($i=1,2$)
\begin{align}
\label{eq:GBGravitonSmear}
& z^2 \langle h_{\mu\nu}(t_1,{\bf x_1},z_1) h_{\mu\nu}(t_2,{\bf x_2},z_2)\rangle(\alpha_1)\nonumber\\
&= {1 \over {\rm vol}(B^d)\left(1-2\alpha_{1}\frac{(d-2)(d-3)}{R^{2}}\right)}
\hspace{-5mm} \int\limits_{\hspace{10mm}t_i'{}^2 + \vert {\bf y}'_i\vert^2 < z_i^2}\hspace{-12mm} \int  \hspace{5mm} dt_i' d^{d-1} y_i' \,
\langle T_{\mu\nu}(t_1 + t_1', {\bf x}_1 + i {\bf y}_1')T_{\mu\nu}(t_2 + t_2', {\bf x}_2 + i {\bf y}_2')\rangle\nonumber\\
& \hbox{\rm where volume of a unit $d$-ball} = {\rm vol}(B^d) = {2 \pi^{d/2} \over d \Gamma(d/2)}
\end{align}

Now if we use
the HKLL dictionary of AdS/CFT in the \emph{holographic gauge},
then we arrive at the boundary CFT stress-tensor two point functions,
\[
\langle T_{\mu\nu}(x)T_{\lambda\rho}(y)\rangle=\lim_{z\rightarrow0}z^{2d}\langle h_{\mu\nu}(z,x)h_{\lambda\rho}(z,y)\rangle.
\]
Thus we see that the stress-tensor two point functions of the $1/\lambda$
corrected CFT is related to the leading $\lambda\rightarrow\infty$
result by,
\begin{equation}
\langle T_{\mu\nu}(x)T_{\lambda\rho}(y)\rangle_\lambda=\frac{1}{\left(1-2\alpha_{1}\frac{(d-2)(d-3)}{R^{2}}\right)}\langle T_{\mu\nu}(x)T_{\lambda\rho}(y)\rangle_{\lambda=\infty}.\label{eq: Stress tensor with 1/lambda corrections in terms of stress tensor at infinite lambda}
\end{equation}
Here $\alpha_1/R^2$ is to be expressed in powers of $\lambda^{-1}$ as in \eqref{eq:alphalambdareln} or in terms of CFT central charges as in \eqref{eq: Coefficient of the Gauss-Bonnet correction to the bulk} below. Further since the central charge is defined to be the coefficient of the leading singularity of the two-point function of the CFT stress
tensor (\ref{eq: Central Charge as the leading singularity of the two point function of stress tensor}), this overall coefficient of the Gauss-Bonnet correction to the bulk determines the `$\lambda$-running' of the central charge as we turn on the marginal coupling $\lambda$\footnote{Similar expressions have also been found in e.g. \cite{Buchel:2009sk} where they consider AdS/CFT for Gauss-Bonnet theory in the bulk from a ``bottom up'' phenomenological approach. Their Gauss-Bonnet term could have contributions from $1/N$ order since the AdS radius after adding the GB term changes compared to the AdS solution in the pure EH gravity. Here we reconstruct the bulk action from the CFT order by order, in an $1/\lambda^\alpha$ expansion.}
\begin{equation}
\frac{\alpha_{1}}{R^{2}}=\frac{1}{2(d-2)(d-3)}\left(1-\frac{C(\lambda\rightarrow\infty)}{C(\lambda)}\right).\label{eq: Coefficient of the Gauss-Bonnet correction to the bulk}
\end{equation}
Above, $C(\lambda)$ denotes the central charge appearing in front of the stress-tensor correlator, but only expanded up to first sub-leading order in $\lambda^{-1}$ expansion. Alternatively, upon defining
\[
C(\lambda)\equiv C_{\infty}+\frac{1}{\lambda^{\alpha}}C^{(1)},
\] 
we can obtain
\begin{equation}\label{eq:altexpforC}
\frac{\alpha_1}{l_{s}^{2}}=\frac{C^{(1)}}{2C_\infty(d-2)(d-3)}.
\end{equation}
This equation predicts that for $d=2,3$, the correction vanishes, $C^{(1)}=0$, which makes sense because in those dimensions the bulk Gauss-Bonnet term either vanishes or is a topological term (not local). Thus knowing the boundary CFT data i.e. the stress tensor two-point
function order by order in $1/\lambda$ expansion, we can determine
the coefficient of the respective Lovelock terms that we need to add in
the gravity lagrangian/action to reconstruct the bulk. Using \eqref{eq: Coefficient of the Gauss-Bonnet correction to the bulk} or \eqref{eq:altexpforC}, the right hand side of \eqref{eq:GBGravitonSmear} becomes a purely boundary quantity.

\section{Conclusion and Outlook}\label{sec:concl}

In this paper we have taken the first step towards incorporating $\frac{1}{\lambda}$ corrections (i.e. finiteness of the marginal coupling in a CFT) in the construction of smeared boundary operators which play the roles of local fields in the AdS bulk. The  construction is performed while holding $N$ infinite i.e. when only connected two point correlators in the CFT are turned on, or equivalently in the limit of Newton's constant $G_N\to 0$ in the AdS bulk, i.e. the bulk theory is at tree-level in quantum gravity. We have studied the effect of the $1/\lambda$ corrections in two cases. In the first case we looked at changes in the bulk theory resulting from  $1/\lambda$ corrections to (two point) correlators of a CFT scalar primary (unprotected). The anomalous dimensions of the CFT primary operators develop a dependence on $\lambda$ and we showed that this leads to new tree-level/classical interactions to the dual bulk AdS scalar theory via non-minimal couplings to the background, i.e. couplings to higher orders of the curvature tensors and scalars. These new interactions can be thought to be arising out of massive string modes since in the context of the gauge-string duality, the $1/\lambda$ corrections are expected to be equivalent to perturbative worldsheet effects ($\alpha'$ corrections). In the second case, we looked at the pure gravity sector in the AdS bulk (dual to the CFT stress-tensor multiplet) and again we found out that CFT $1/
\lambda$ corrections to the $TT$ correlators transpire into higher curvature correction terms but only those which are of the special Lovelock form. We have thus shown that the (HKLL) map from local bulk operators to non-local boundary operators via the smearing functions can be easily and very naturally extended from the $\lambda\to \infty$ case to include $1/\lambda$ corrections (equations \eqref{eq:bulkhcsmear} and \eqref{eq:GBGravitonSmear}).


There are various avenues for future directions. Our work is only the first step (the next to leading order in $1/\lambda$) towards understanding the emergence of higher curvature and higher derivative terms on the bulk, i.e. stringy physics from the underlying CFT. The logical next step would be to extend our results to incorporate higher order $1/\lambda$ corrections. It remains to be seen, how does the standard HKLL AdS/CFT bulk-boundary map morph. Another natural generalization would be to consider both the scalars and the gravitational degrees of freedom interacting at a certain order of $\alpha'$. Our results seem to suggest that as long as we have full AdS isometry, the scalar and metric smearing functions will again have the same structures, however the overall normalization factors and powers of the AdS covariant distance function $\sigma(x,x')$ will change. The AdS isometry considerations dictate that it is also straight-forward to generalize our situation for asymptotically AdS spacetimes which are global quotients of pure AdS, viz. AdS$_3$-Rindler, BTZ and higher dimensional hyperbolic black holes. We expect our results to be easily adapted to go through for such backgrounds. 

However one of the main goal of our studies of bulk locality (micro-causality) from boundary is to better understand the AdS quantum gravity itself and perhaps precisely derive the restrictions placed on the dual CFT in order to have a local causal bulk AdS. Along this line one of the recent interesting developement was in \cite{Camanho:2014apa}, which pointed out the need of an infinite tower of higher spin particles in the AdS bulk in order to construct a consistent (causality respecting) theory of quantum gravity. In AdS/CFT, one should see such a structure purely from the dual CFT. In \cite{Camanho:2014apa} the authors found their results by looking at bulk three-point graviton vertex, or equivalently in the CFT stress tensor three-point function. This is an effect which is mainfest at an order subleading in $1/N$. As we mentioned in the introduction, within the HKLL programme itself, already a lot of literature exist that deal with $1/N$ corrections to fields of various spins such as spin 2 \cite{Kabat:2013wga} and higher \cite{Sarkar:2014dma}, and thus the next step will be to understand such effects simultaneously with our $\frac{1}{\lambda}$ corrections. However, we believe that to derive the results of \cite{Camanho:2014apa} would require us to consider another criterion in addition to micro-causality in the bulk. Micro-causality is a feature of local quantum field theories (in this curved space QFT) and it guarantees causal propagation of information. However, this is not expected to be enough when one considers stringy physics, which is not described by a local QFT lagrangian i.e. containing only a finite number of terms. For causal propagation in such stringy physics one requires additional consistency/causality criterion in the bulk. The criterion of a gravitational Shapiro delay used by \cite{Camanho:2014apa} could exactly be such a constraint on a non-local yet causal bulk theory. Another standard expectation that comes from the studies of these massive stringy modes is that for a sub-AdS duality one needs to have a gap in conformal dimensions between fields of spin 2 and higher \cite{Heemskerk:2009pn}. In fact, in \cite{Afkhami-Jeddi:2016ntf} it was shown that a key result of \cite{Camanho:2014apa} (eq. 5.20 of \cite{Camanho:2014apa}) can be rederived from a CFT by \emph{assuming} this gap and assuming the chaos bound on out of time ordered four point functions of \cite{Maldacena:2015waa}. So far our prescriptions, depending solely on bulk micro-causality, is insensitive to these extra constraints. However, from our point of view i.e. from the point of \emph{ab initio} reconstructing the bulk from the CFT, these conditions or restrictions should emerge naturally from the existence of a perturbative expansion of OPE's in two parametrically large dimensionless quantities, namely $N$ and $\lambda$. It would be nice to see the emergence of a gap in the spectrum which is related to the marginal coupling of the CFT (without using a bulk stringy spectrum).

\hfill{}\rule[0.3ex]{0.6\columnwidth}{1pt}\hfill{}

\bigskip{}
\centerline{\textbf{Acknowledgements}}

We thank Dan Kabat for helpful discussions and for giving his valuable feedback
on the draft. SR thanks Justin David  and Aninda Sinha for enlightening discussions and Chethan Krishnan for pointing out reference \cite{Mazzucato:2009fv}. Special thanks are due to Arpan Bhattacharyya for pointing out many references in higher derivative gravity literature, and to
Kallol Sen for telling us about their interesting work in \cite{Sen:2014nfa}. SR also wishes to thank Prof. Shibaji Roy for their hospitality at the Saha Institute of Nuclear Physics (SINP), Kolkata during summer 2014, where part of this work was being conducted.  SR is supported by IIT Hyderabad seed grant SG/IITH/F171/2016-17/SG-47. The work of SR in Israel was partially supported by the American-Israeli Bi-National Science Foundation, the Israel Science Foundation Center of Excellence and the I-Core Program of the Planning and Budgeting Committee, and the Israel Science Foundation's ``The Quantum Universe". The work of DS is funded by the ERC grant `Selfcompletion'.\\

\appendix
\noindent 

\section{Gravitational perturbation results\label{sec:Gravitational-Perturbation}}

The form of the gravitational perturbation theory used in this paper%
\footnote{In metric perturbation theory it is customary to multiply the perturbation
by $\kappa=\sqrt{8\pi G}$,
\[
g_{AB}=g_{AB}^{(0)}+\kappa\delta g_{AB}.
\]
However we will not follow this convention in this paper.%
}

\begin{eqnarray*}
g_{AB} & = & g_{AB}^{(0)}+\delta g_{AB},\\
g^{AB} & = & g^{(0)AB}-\delta g^{AB}+\delta g^{AL}\delta g_{L}^{B}+\ldots,\\
\mathcal{R}_{AB} & = & \mathcal{R}_{AB}^{(0)}+\Delta^{(1)}\mathcal{R}_{AB}+\Delta^{(2)}\mathcal{R}_{AB}+\ldots,\\
\Delta^{(1)}\mathcal{R}_{AB} & = & \nabla_{(A}\nabla^{C}\delta g_{B)C}-\frac{1}{2}\square\delta g_{AB}-\frac{1}{2}\nabla_{B}\nabla_{A}\delta g_{C}^{C}+\mathcal{R}_{AC}^{(0)}\delta g_{B}^{C}-\mathcal{R}_{ACBD}^{(0)}\delta g^{CD}\\
\Delta^{(2)}\mathcal{R}_{AB}g^{(0)AB} & = & \delta g_{AB}\left(\frac{g^{(0)AB}g^{(0)CD}-g^{(0)AC}g^{(0)BD}}{4}\square\right.\\
 &  & \left.+\frac{\nabla^{A}\nabla^{C}g^{(0)BD}-g^{(0)CD}\nabla^{A}\nabla^{B}}{2}+\frac{\mathcal{R}^{(0)AC}g^{(0)BD}-\mathcal{R}^{(0)ACBD}}{2}\right)\delta g_{CD}
\end{eqnarray*}
\begin{eqnarray*}
\Delta^{(1)}\mathcal{R}_{AB}\delta g^{AB} & = & \delta g_{AB}\left(\nabla^{A}\nabla^{C}g^{(0)BD}-\frac{1}{2}g^{(0)AC}g^{(0)BD}\square-\frac{1}{2}\nabla^{A}\nabla^{B}g^{(0)CD}\right.\\
 &  & \left.+\mathcal{R}^{(0)AC}g^{(0)BD}-\mathcal{R}^{(0)ACBD}\right)\delta g_{CD}
\end{eqnarray*}
\[
\Delta^{(2)}\mathcal{R}=\frac{1}{4}\left(g^{(0)AB}g^{(0)CD}+g^{(0)AC}g^{(0)BD}\right)\square+\frac{1}{2}\nabla^{A}\nabla^{C}g^{(0)BD}+\frac{1}{2}\left(\mathcal{R}^{(0)AC}g^{(0)BD}+\mathcal{R}^{(0)ACBD}\right)
\]

\subsection{First order perturbation expressions}

We have used the following first order perturbations of gravitational
quantites\footnote{A collection of relevant formulas can also be found in \cite{Deser:2002jk} e.g.}

\begin{eqnarray*}
\delta\Gamma_{\nu\sigma}^{\mu} & = & \frac{1}{2}g^{\mu\alpha}\left(\nabla_{\nu}\delta g_{\alpha\sigma}+\nabla_{\sigma}\delta g_{\nu\alpha}-\nabla_{\alpha}\delta g_{\nu\sigma}\right),\\
\delta \mathcal{R}^{\sigma}\,_{\rho\mu\nu} & = & \nabla_{\mu}\delta\Gamma_{\nu\rho}^{\sigma}-\nabla_{\nu}\delta\Gamma_{\mu\rho}^{\sigma},\\
 & = & \frac{1}{2}\left(\nabla_{\mu}\nabla_{\nu}\delta g_{\rho}^{\sigma}+\nabla_{\mu}\nabla_{\rho}\delta g_{\nu}^{\sigma}-\nabla_{\mu}\nabla^{\sigma}\delta g_{\rho\nu}-\mu\rightarrow\nu\right).\\
\delta \mathcal{R}_{\mu\nu} & = & \nabla_{\rho}\delta\Gamma_{\mu\nu}^{\rho}-\nabla_{\nu}\delta\Gamma_{\rho\mu}^{\rho}\\
 & = & \frac{1}{2}\left(\nabla^{\sigma}\nabla_{\mu}\delta g_{\sigma\nu}+\nabla^{\sigma}\nabla_{\nu}\delta g_{\mu\sigma}-\square\delta g_{\mu\nu}-g^{\rho\sigma}\nabla_{\nu}\nabla_{\mu}\delta g_{\rho\sigma}\right),\\
\delta \mathcal{R} & = & \nabla^{\mu}\nabla^{\nu}\delta g_{\mu\nu}-g^{\mu\nu}\square\delta g_{\mu\nu}-R^{\mu\nu}\delta g_{\mu\nu},\\
\delta\sqrt{-g} & = & \frac{1}{2}\sqrt{-g}g^{\mu\nu}\delta g_{\mu\nu}
\end{eqnarray*}
One also needs the following results using the Bianchi Identity,%
\begin{eqnarray*}
\nabla^{A}\mathcal{R}_{AB} & = & \frac{1}{2}\nabla_{B}\mathcal{R}\\
\nabla^{A}\nabla^{B}\mathcal{R}_{AB} & = & \frac{1}{2}\square \mathcal{R}
\end{eqnarray*}
and %
\begin{eqnarray*}
\nabla^{M}\nabla^{L}\mathcal{R}_{ALBM} & = & \square \mathcal{R}_{AB}-\nabla^{M}\nabla_{B}\mathcal{R}_{AM}\\
 & = & \square \mathcal{R}_{AB}-\nabla_{B}\nabla_{M}\mathcal{R}_{A}\,^{M}-\left[\nabla_{M},\nabla_{B}\right]\mathcal{R}_{A}\,^{M}\\
 & = & \square \mathcal{R}_{AB}-\frac{1}{2}\nabla_{A}\nabla_{B}\mathcal{R}-\mathcal{R}_{AC}\mathcal{R}_{B}\,^{D}+\mathcal{R}_{ACBD}\mathcal{R}^{CD}
\end{eqnarray*}

\rule[0.5ex]{1\columnwidth}{1pt}


\bibliographystyle{brownphys}
\bibliography{lambda}

\end{document}